\documentstyle[prb,aps,amsfonts,amssymb,epsf,multicol]{revtex}

\begin{document}
\draft 
\preprint{{\bf ETH-TH/98-??}}

\title{Casimir Force between Vortex Matter in Anisotropic and Layered
Superconductors}

\author{H.~G.~Katzgraber$^{a,b}$, H.~P.~B\"uchler$^{a}$, 
and G.~Blatter$^{a}$}

\address{$^a$Theoretische Physik, ETH-H\"onggerberg, CH-8093
  Z\"urich, Switzerland}

\address{$^b$Physics Department, University of California, Santa Cruz
  CA 95064, USA}

\date{\today}
\maketitle
\begin{abstract}
  
  {We present a new approach to calculate the attractive 
   long range vortex-vortex interaction of the van der Waals 
   type present in anisotropic and layered superconductors. The
   mapping of the statistical mechanics of vortex lines onto the
   imaginary time quantum mechanics of two dimensional charged bosons
   allows us to define a 2D Casimir problem: Two half-spaces of
   (dilute) vortex matter separated by a gap of width $R$ are mapped
   to two dielectric half-planes of charged bosons interacting via a
   massive gauge field.  We determine the attractive Casimir force
   between the two half-planes and show, that it agrees with the
   pairwise summation of the van der Waals force between vortices
   previously found by Blatter and Geshkenbein [Phys.~Rev.~Lett.~{\bf
   77}, 4958 (1996)].}

\end{abstract}


\begin{multicols}{2}
\narrowtext
\vspace*{-1.5truecm} 

\section{Introduction} 

Within the Shubnikov phase of type II superconductors the applied
magnetic field enters the sample in the form of flux lines.  The
standard mean-field type calculation\cite{Tinkham} shows that in an
isotropic material two straight vortices repel at all distance scales,
with an interaction strength $V_{\rm rep} = 2 \varepsilon_0
K_0(R/\lambda)$, $\varepsilon_0 = (\Phi_0/4\pi\lambda)^2$ being the
basic energy scale in the vortex matter, $K_0$ is the zero-order
modified Bessel function, $R$ is the inter-vortex distance, and
$\lambda$ the magnetic penetration length ($\Phi_0 = hc/2e$ denotes
the flux quantum). However, it has recently been shown \cite{GB} that
in layered and strongly anisotropic superconductors the thermal
fluctuations of the flux lines give rise to a long range attraction
$V_{\rm vdW} \sim - (T/d) (\lambda/R)^4$ of the van der Waals type
between the vortices, where $T$ denotes the temperature and $d$ is the
interlayer separation.  The strongly fluctuating and layered high
temperature superconductors are particularly well suited to exhibit
this attractive component in the vortex-vortex interaction.
Alternatively, the attraction is induced through static vortex
distortions due to an underlying pinning landscape, an effect recently
studied by Mukherji and Nattermann\cite{MN} and by Volmer {\it et
  al.}\cite{VMN}

Following a suggestion of Nelson\cite{Nelson}, the statistical
mechanics of vortices can be mapped to the imaginary time quantum
mechanics of two-dimensional (2D) bosons.  The particular interaction
between the flux lines renders the bosons charged (with a charge
screened on the scale of the London penetration depth $\lambda$). This
type of long range interaction can be formulated in terms of a massive
gauge field theory\cite{popov,Feigelman,RMP}. Within the resulting 2D
massive electrodynamics, the vortex matter acts as a dielectric medium
and we can define a Casimir problem, see Fig.~1: Under the vortex
$\rightarrow$ boson mapping two half-spaces of vortices separated by a
gap of width $R$ act as two dielectric planar media which attract each
other via a Casimir force.
\begin{figure}
\centerline{\epsfxsize= 8.5cm\epsfbox{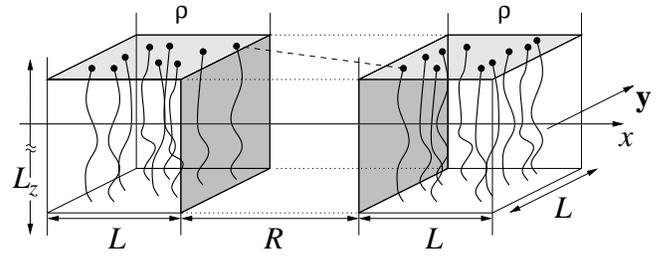}}
\vspace{0.5cm}
\caption{Geometry used in the calculation of the Casimir force between
  two half spaces of vortex matter separated by a gap of width $R$.
  Shown are two parallel cubes of size $L_z\times L^2$ at a distance
  $R$ with vortex density $\rho$ ($L_z \rightarrow \infty$).  We
  choose a coordinate system with the $z$-axis parallel to the 
  vortex lines and the $x$-axis perpendicular to the 
  planes of the gap (dark shaded planes).  The statistical
  mechanics of the flux lines is mapped to the imaginary time quantum
  mechanics of two dimensional (2D) bosons (solid circles).  The cut
  through the vortex lines (lightly shaded planes) defines a time
  slice of the 2D bosons. The Casimir energy produced by the
  dielectric action of the two media is calculated by summing over the
  ground state energies of the fluctuating massive electromagnetic
  field mediating the interaction between the 2D bosons.  Using the
  technique of pairwise summation (dashed line) valid for rarefied
  media (i.e., $\rho \rightarrow 0$), we can relate the macroscopic
  Casimir effect with the microscopic van der Waals attraction between
  the particles involved. Picture not to scale.}
\label{fig:1}
\end{figure}
In the present work, we determine the Casimir force between two
dielectric half planes of charged bosons. For dilute media, the
macroscopic Casimir force can be related to the microscopic van der
Waals force between the media's constituents via pairwise summation.
Here, we present a derivation of the van der Waals force between
vortex lines via this alternative route, calculating first the Casimir
force between two dielectric planar media in the boson picture and
then reconstructing the van der Waals force in the reductionist way.

Since the discovery of the Casimir effect\cite{Casim} in 1948, several
hundred papers have dealt with this phenomenon, disseminating its
fascination into many branches of physics
\cite{fbr1,fbr2,fbr3,fbr4,fbr5,fbr6,fbr7,Mostepanenko}. Casimir forces
between macroscopic bodies are a quantum effect caused by a shift in
the zero-point energy of gauge field fluctuations such as the
electromagnetic one. The Casimir effect is bound to a number of system
properties, such as topology and dielectric permittivity, and the
reduction to an analogous van der Waals interaction is not always
possible. However, the interpretation in terms of a van der Waals
attraction is possible for the case of rarefied media and appropriate
geometries, such as the parallel plate setup\cite{Mostepanenko}. In
our derivation of the van der Waals force from the Casimir effect we
will make use of these special conditions.

In the following, we discuss the relationship between the Casimir
effect and the van der Waals interaction within a path integral
formulation (Sec.~II) before deriving the appropriate action for the
2D charged bosons from the London functional describing the vortices
(Sec.~III). In section IV, we briefly review the derivation of the van
der Waals force in the original vortex language and then proceed with
the calculation of the 2D Casimir effect in section V, the main
section of the paper containing the new results.

\section{Casimir versus van der Waals} 

We consider two parallel material slabs made from fluctuating dipoles
and separated by a vacuum gap. Summing pairwise over all microscopic
van der Waals interactions between the dipoles provides the
macroscopic Casimir interaction between the slabs. On the other hand,
we can determine the dielectric properties of the individual slabs as
produced by the fluctuating dipoles. The specific boundary conditions
due to the dielectric properties of the slabs influence the spectrum
of the electromagnetic field confined in between. The change in the
spectrum as a function of the separation of the slabs produces the
Casimir force.  This analogy between the Casimir and the van der Waals
force is transparently brought out within a path integral formulation,
see Fig.~2: Assume the system under consideration can be described by
an action ${\cal S}[{\bf a}, {\bf j}]$ depending on the gauge field
${\bf a}$ and a particle current ${\bf j}$.  Carrying out the partial
integration in the partition function ${\cal Z} = \int {\cal D}[{\bf
  a}]{\cal D}[{\bf j}]\, e^{-{\cal S} [{\bf j},{\bf a}]/\hbar}$ over
the matter field ${\bf j}$ or over the gauge field ${\bf a}$, we
obtain an effective action ${\cal S}_{\rm eff}$ describing the
conjugate field alone: From the effective action ${\cal S}_{\rm
  eff}[{\bf a}]$ describing the gauge field we can derive the Casimir
effect, while the interaction of the particle currents as described by
${\cal S}_{\rm eff}[{\bf j}]$ will give us the van der Waals
attraction.

The Casimir- and van der Waals forces then can be related to one
another via the pairwise summation of the interparticle
forces\cite{Israel}: Consider two $d$-dimensional homogeneous
macroscopic dielectric bodies of density $\rho$ with parallel
interfaces separated by a distance $R$. The two bodies attract one
another due to a microscopic particle-particle interaction $V_{\rm
  vdW} = \Lambda \: r^{-\alpha}$, $\alpha > d$, of the van der Waals
type. The interaction energy can be written in the form
\[
   U(R) = L^{d-1} \rho^2 \int_{R}^{\infty} \!\! {\rm d}x
          \int_x^{\infty} \!\! {\rm d}x' \int_{E}{\rm d}^{d-1}y 
          \, V_{\rm vdW}(r),
\]
where $E$ is a hypercube of size $L^{d-1}$ parallel to the interface
and $r = \sqrt{x^2 + y^2}$, with ${\bf y}$ the $d-1$ dimensional
in-plane coordinate, while $x$ is the coordinate along the direction
perpendicular to the plane. We then find for the Casimir force
density\cite{Lang} $f = - \partial_R U(R)/L^{d-1}$ the result
\begin{eqnarray}
   f &=& \rho^2 \int_{R}^{\infty}\!{\rm d}x\int_{E}{\rm d}^{d-1}y 
                \, V_{\rm vdW}(r) \nonumber \\ &=&
                \Lambda \rho^2 \: \pi^{(d-1)/2}
                \frac{\Gamma[(1+\alpha-d)/2]} {\Gamma(\alpha/2)}
                \frac{1}{\alpha-d} \frac{1}{R^{\alpha-d}}.
\label{F_of_R}
\end{eqnarray}
The result (\ref{F_of_R}) then allows to infer the parameters
$\Lambda$ and $\alpha$, characterizing the van der Waals interaction
$V_{\rm vdW}$, from the macroscopic Casimir force density $f$.
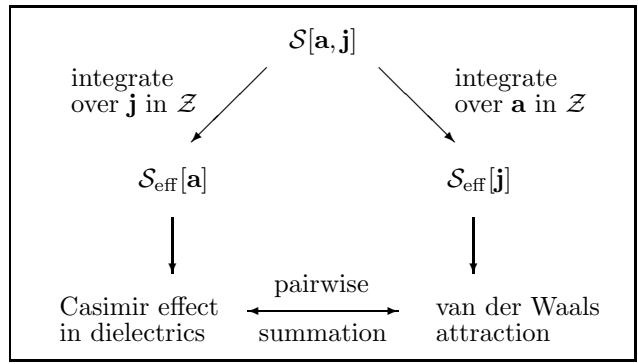
\begin{figure}
\begin{center}
\setlength{\unitlength}{1cm}
\begin{picture}(8.5,5)
\put(0.1,0.1){\framebox(8.3,4.7){}}
\put(2.25,2){\vector(0,-1){0.75}}
\put(6.25,2){\vector(0,-1){0.75}}
\put(4.25,0.75){\vector(-1,0){1}}
\put(4.25,0.75){\vector(1,0){1}}
\put(3.5,4){\vector(-1,-1){1}}
\put(5,4){\vector(1,-1){1}}
\put(3.8,4.25){${\cal S}[{\bf a}, {\bf j}]$}
\put(1.8,2.4){${\cal S}_{\rm eff}[{\bf a}]$}
\put(5.9,2.4){${\cal S}_{\rm eff}[{\bf j}]$}
\put(0.75,0.7){Casimir effect}
\put(0.75,0.35){in dielectrics}
\put(5.75,0.7){van der Waals}
\put(5.75,0.35){attraction}
\put(3.4,0.35){summation}
\put(3.6,1){pairwise}
\put(0.9,3.75){integrate}
\put(0.9,3.4){over ${\bf j}$ in ${\cal Z}$}
\put(6,3.75){integrate }
\put(6,3.4){over ${\bf a}$ in ${\cal Z}$}
\end{picture}
\vspace{0.5cm}
\caption{Schematic relation between the Casimir effect and the 
van der Waals attraction.}  
\label{Fig:2} 
\end{center}
\end{figure}

After mapping the thermally fluctuating vortex matter to a system of
2D quantum charged bosons, we will arrive at an action ${\cal S}[{\bf
  j}, {\bf A}, {\bf a}]$ with two gauge fields ${\bf a}$ and ${\bf
  A}$, see Eq.~(\ref{edr}).  The Casimir effect we are interested in
here is the one produced by the fake gauge field ${\bf a}$: The
integration over the matter field ${\bf j}$ will produce the
dielectric properties of the media, while the integration over the
physical gauge field ${\bf A}$ renders the fake field massive and
hence will always exponentially confine the Casimir force to finite
distances.

\section{Boson Representation of Vortices}

We start from the London free energy in an isotropic
superconductor\cite{Tinkham}
\begin{equation}
{\cal F}[{\bf B}] = \frac{1}{8\pi}\int {\rm d}^3r \, [{\bf B}^2 +
\lambda^2(\nabla \times {\bf B})^2],
\label{London}
\end{equation}
with ${\bf B} = \nabla \times {\bf A}$ the magnetic field and
$\lambda$ denoting the London penetration depth. In order to acquaint
for the vortices, we add the current term $-({\bf j}\cdot{\bf B})
\Phi_0/8\pi$ with
\begin{equation}
{\bf j} = ({\bf J},\rho) = \sum_\mu {\bf t}_\mu(z) \,
\delta^{(2)}({\bf R} - {\bf R}_\mu (z)),
\label{current}
\end{equation}
where ${\bf r} = ({\bf R},z)$, ${\bf R} = (x,y) \in {\mathbb{R}}^2$,
the coordinates ${\bf R}_\mu (z)$ denote the position, the vectors
${\bf t}_\mu = (\partial_z {\bf R}_\mu (z),1)$ the direction of the
vortex lines, and $\Phi_0 = hc/2e$ is the unit of flux.  Ignoring
screening, the interaction between the vortex lines is long ranged and
thus can conveniently be expressed through a mediating gauge field
${\bf a}$. As usual, we introduce the gauge field ${\bf a}$ as an
auxiliary field such that $\int {\cal D}{\bf a} \exp(-\beta {\cal
  F}'[{\bf a}]) = \exp(-\beta {\cal F})$, with $\beta = 1/T$ the
inverse temperature\cite{RMP} (we set the Boltzmann constant
$k_{\scriptscriptstyle B}$ to unity and fix the gauge through the
condition $\nabla \cdot {\bf a} = 0$),
\begin{eqnarray} 
{\cal F}'[{\bf a}, {\bf A}, {\bf j}] &=& \int {\rm d}^3{x} \,
\left[i{\bf a}\cdot\left({\bf j}-\frac{1}{\Phi_0}(\nabla\times{\bf
A})\right) \right.
\label{Fprime} \\
&& \quad + \left. \frac{1}{8\pi}(\nabla \times {\bf A})^2 +
\frac{1}{2g^2}(\nabla \times {\bf a})^2 \right], \nonumber
\end{eqnarray}
where $g^2 = 4\pi \varepsilon_0$ and the line energy $\varepsilon_0 =
(\Phi_0/4\pi \lambda)^2$ is the basic energy scale in the problem.  In
(\ref{Fprime}), we have accounted for screening by introducing back
into the model the real gauge field ${\bf A}$.

The anisotropy of uniaxial layered materials is most conveniently
introduced through a rescaling of the scalar fake magnetic field $b =
(\nabla \times {\bf a})_z$ \cite{BGL,BGvdWpC},
\[
(\nabla\times{\bf a})_z\rightarrow
\frac{1}{\varepsilon}(\nabla\times{\bf a})_z,
\]
where the anisotropy factor $\varepsilon = \sqrt{m_{ab}/m_c}$ is
determined through the effective masses perpendicular ($m_c$) and
parallel ($m_{ab}$) to the $ab$-plane (the fake electric field ${\bf
  e}_\perp = (e_y, -e_x) = -(\nabla \times {\bf a})_{xy}$ remains
unchanged; the subscript `$xy$' identifies the planar component of a
vector, ${\bf a} = ({\bf a}_{xy}, a_z)$).

We map the statistical mechanics of the vortex system to the imaginary
time quantum mechanics of 2D bosons through the
replacements\cite{Nelson,Feigelman,RMP} $z \rightarrow \tau$
(imaginary boson time), $\beta^{-1} \rightarrow
\hbar^{\scriptscriptstyle B}$ (the boson's Planck constant), ${\cal
  F}' \rightarrow {\cal S}$ (the boson action), and $1/\varepsilon
\rightarrow c$ (the light velocity in the boson system). The boson
partition function is given by
\begin{equation}
{\cal Z} = \int {\cal D}[{\{\bf R}_\mu\}] {\cal D}[{\bf A}]{\cal
D}[{\bf a}] e^{-{\cal S}[\{{\bf R}_\mu\},{\bf A},{\bf
a}]/\hbar^{\scriptscriptstyle B}},
\label{zzz}
\end{equation}
with ${\cal S} = {\cal S}_0 + {\cal S}_{\rm int}$ and
\begin{eqnarray}
&& {\cal S}_0 [\{{\bf R}_\mu\}] =
\int \!\! {\rm d}\tau \sum_\mu \left[\frac{m}{2}
\left(\partial_\tau {\bf R}_\mu (\tau)\right)^2 -
\mu^{\scriptscriptstyle B} \right],\label{plm}\\
&&{\cal S}_{\rm int} [\{{\bf R}_\mu\},{\bf A},{\bf a}]
= \int \!\! {\rm d} \tau {\rm d}^2 R \left[i{\bf a}\cdot
\left({\bf j}-\frac{1}{\Phi_0} (\nabla \times {\bf A})\right)
\right. \label{edr} \\
&& \quad + \left. \frac{1}{8\pi} (\nabla \times {\bf A})^2
+\frac{1}{2g^2}\left((\nabla\times{\bf a})_{xy}^2 + 
\frac{1}{\varepsilon^2} (\nabla\times {\bf a})_\tau^2\right)
\right].
\nonumber
\end{eqnarray}
Here, $\mu^{\scriptscriptstyle
  B}=H\Phi_0/4\pi-\varepsilon_0\ln(\lambda/\xi)$ is the chemical
potential of the bosons, $\rho = B/\Phi_0$ is the boson density, and
$m = \varepsilon_0$ is the boson mass (we have introduced a term
$-{\bf B}{\bf H}/4\pi$ to account for the external magnetic field
${\bf H}$ producing the vortices in the superconductor; $\xi$ denotes
the planar coherence length). In the thermodynamic limit, we have $L_z
\rightarrow \infty$, corresponding to $T^{\scriptscriptstyle B} =
\hbar^{\scriptscriptstyle B} / L_z \rightarrow 0$, i.e., we are
interested in the ground state physics of the boson system.  Note that
the free boson action ${\cal S}_0$ contains the bare mass $m$ due to
the vortex core energy. The retarded self-interaction of the bosons
via their gauge fields then produces the mass renormalization $m
\rightarrow m^{\scriptscriptstyle B} = \varepsilon_l(k_z)$, where
$\varepsilon_l(k_z)$ is the dispersive line tension\cite{RMP} of the
vortex lines.

\section{Van der Waals Interaction}

To set the stage, we briefly review the derivation of the van der
Waals interaction as presented in Ref.\ \onlinecite{GB}; this will
allow us to fix some flaws in the previous derivation and will provide
us with a check on the results for the Casimir force derived later.

For a simple qualitative analysis we consider two vortices in a
layered superconductor (with layers separated by the distance $d$).
Ignoring the coupling between the layers, the fluctuating pancake
vortices interact via a logarithmic potential $V(R) \approx 2
\epsilon_0 \ln(R/\lambda)$. Second order perturbation theory then
provides us with a van der Waals interaction $-V_{\rm vdW} \sim (T/d)
(\lambda/R)^4$, the energy scale $T/d$ being set by the driving action
of the thermal fluctuations.  At long distances $R > d/\varepsilon$ a
finite interlayer coupling changes this result as the Josephson
interaction reduces the cutoff $1/d$ on the $k_z$ modes to the new
value $1/\varepsilon R$; the van der Waals interaction crosses over to
$V_{\rm vdW} \sim -(T/\varepsilon\lambda) (\lambda/R)^{-5}$.  The two
results have their analogue in the van der Waals attraction between
neutral atoms, where the interaction potential exhibits a crossover
from a $r^{-6}$ at short to a $r^{-7}$ behavior at large
distances\cite{LL}.

We proceed with the derivation of the long range van der Waals
interaction between vortex lines/bosons. Following the scheme in Fig.\ 
2, we integrate over the gauge fields ${\bf A}$ and ${\bf a}$ in the
partition function (\ref{zzz}) and obtain the effective
current-current interaction
\begin{equation}
{\cal F} [{\bf j}] = 
\frac{\varepsilon_0}{2}\int {\rm d}^3r \,{\rm d}^3 r'\, 
{j}_\alpha({\bf r})V_{\alpha\beta}^{\rm int}({\bf r}-{\bf r}')
{j}_\beta({\bf r})
\label{aa1}
\end{equation}
(we return to the more natural statistical mechanics notation in this
section, ${\cal S}_{\rm eff} \rightarrow {\cal F}$).  Inserting the
expression (\ref{current}) for the currents, we can cast
Eq.~(\ref{aa1}) into the standard form\cite{RMP}
\begin{equation} 
{\cal F}[\{{\bf r}_\mu\}] = \frac{\varepsilon_0}{2} \sum_{\mu,\nu}
\int {\rm d}r_{\mu\alpha} {\rm d}r'_{\nu\beta} 
V_{\alpha\beta}^{\rm int} ({\bf r}_\nu - {\bf r}'_\mu),
\label{aa2}
\end{equation}
with the vortex positions ${\bf r}_\mu = ({\bf R}_\mu,z)$ and the
interaction potential $V_{\alpha\beta}^{\rm int}$, conveniently
expressed within a Fourier representation,
\begin{equation}
V_{\alpha\beta}^{\rm int}({\bf r}-{\bf r}') = 4\pi \lambda^2 \int
\frac{{\rm d}^3{k}}{(2\pi)^3} e^{i{\bf k}({\bf r}-{\bf r}')}
V_{\alpha\beta}^{\rm int}({\bf k}),
\end{equation}
with 
\begin{eqnarray}
V_{\alpha \beta}^{\rm int}({\bf k})=&& \frac{e^{-(\xi^2 K^2 +\xi_c^2
k_z^2)}} {1+\lambda^2k^2} \nonumber \\ && \times \left[\delta_{\alpha
\beta} - \frac{(\lambda_c^2 -\lambda^2) K_{\perp\, \alpha}K_{\perp\,
\beta}}{1+\lambda^2 k^2 + (\lambda_c^2 -\lambda^2)K^2}\right].
\label{Vab}
\end{eqnarray}
Here ${\bf k} = ({\bf K},k_z) = (k_x, k_y, k_z)$, ${\bf K}_\perp =
(k_y, -k_x)$ and $\lambda_c = \lambda/\varepsilon$, $\xi_c =
\varepsilon \xi$. The free energy (\ref{aa2}) can be split into the
self energy part ${\cal F}^{\, 0}$ ($\mu = \nu$) and the interaction
part ${\cal F}^{\,\rm int}$ ($\mu \neq \nu$). Restricting ourselves to
two vortices a distance $R$ apart, we obtain
\begin{eqnarray}
{\cal F}^{\,\rm int} &=& \frac{\Phi_0^2}{4\pi} 
\int \frac{{\rm d}^3{k}}{(2\pi)^3}dz_{1}dz_{2} \,
e^{i{\bf K}[{\bf R} + {\bf u}_1-{\bf u}_2]}e^{ik_z(z_1-z_2)}
\nonumber \\ &\times&
[V_{zz}^{\rm int}({\bf k}) + t_{1\alpha}(z_1)t_{2\beta}(z_2)
V_{\alpha \beta}^{\rm int}({\bf k})], 
\end{eqnarray}
where we have split the vortex positions $\{{\bf R}_1,{\bf R}_2\}$
into a mean field part $\{{\bf R}^0_1,{\bf R}^0_2\} = \{{\bf 0},{\bf
  R}\}$ and a fluctuating part $\{{\bf u}_1,{\bf u}_2\}$. Up to a
constant, the free energy is given by
\begin{equation}
F(R) = -T\ln Z(R) = -T \ln \langle \exp [-\beta{\cal F^{\rm
int}}]\rangle_0. 
\end{equation}
The average $\langle \dots \rangle_0$ has to be taken with respect to
the self-energy ${\cal F}^{\, 0}$ of the free vortices. Performing a
cumulant expansion, we obtain the effective vortex-vortex interaction
in the form $L_z V_{\rm eff} \approx \langle{\cal F}^{\,\rm int}
\rangle_0 - [\langle ({\cal F}^{\,\rm int})^2 \rangle_0 -\langle {\cal
  F}^{\,\rm int}\rangle_0^2]/2T$, where $L_z$ denotes the sample
thickness. Splitting into longitudinal (to the induction, ${\cal
  F}_\parallel$) and transverse (${\cal F}_\perp$) parts, the
longitudinal interaction produces the standard repulsive vortex-vortex
interaction\cite{Tinkham}, to lowest order in ${\bf u}_\mu$,
\begin{equation}
V_{\rm rep}(R) = 2 \varepsilon_0 K_0(R/\lambda),
\label{rep_eq}
\end{equation}
while higher orders in ${\bf u}_\mu$ merely renormalize the
pre\-factor.  The transverse part produces the van der Waals
interaction\cite{GB}
\begin{equation}
V_{\rm vdW}=-\frac{\langle{\cal F}_\perp{\cal F}_\perp\rangle_0}{2TL_z}.
\label{vdW_def}
\end{equation}
Using the decomposition
\begin{eqnarray}
&&\langle t_{1\alpha}(-k_z) t_{2\beta}(k_z) t_{1\alpha'}(-k_z') 
t_{2\beta'}(k_z') \rangle_0 \nonumber \\
&&= 2 \pi L_z \delta(k_z-k_z') \delta_{\alpha\alpha'} 
\delta_{\beta\beta'} \frac{\langle t^2(k_z)\rangle_0^2}{4}, 
\nonumber
\end{eqnarray}
we can reduce the average in (\ref{vdW_def}) to the simpler form
\begin{equation}
\langle {\cal F}_\perp {\cal F}_\perp\rangle_0 = \frac{\Phi_0^2 L_z}
{64 \pi^2}\int \frac{{\rm d}k_z}{2\pi} [V^{\rm int}_{\alpha\beta}
({\bf R},k_z)]^2 \langle t^2(k_z) \rangle_0^2,
\end{equation}
with the partial Fourier transform
\begin{equation}
V^{\rm int}_{\alpha\beta}({\bf R},k_z) = \int \frac{{\rm d}^2{
K}}{(2\pi)^2}V^{\rm int}_{\alpha\beta}({\bf K},k_z)e^{i{\bf K}{\bf
R}}.
\label{Vpartial}
\end{equation}
In strongly anisotropic and layered material the single vortex mean
squared amplitude of fluctuations $\langle t^2(k_z)\rangle_0^2$ is
limited by the electromagnetic interaction through the dispersive
elasticity $\varepsilon_l(k_z)$,
\begin{equation}
\langle t^2(k_z)\rangle_0 = \frac{2 T}{\varepsilon_l(k_z)},
\label{t^2}
\end{equation}
with
\begin{equation}
\varepsilon_l(k_z)=\frac{\varepsilon_0}{2\lambda^2k_z^2} 
\ln(1+\lambda^2k_z^2). 
\label{elasticity}
\end{equation}
The evaluation of the partial Fourier transform (\ref{Vpartial}) is
carried out in Appendix A and making use of the results
(\ref{app_a_x}) and (\ref{app_a_y}) in the limit $\varepsilon = 0$, we
find the van der Waals interaction in the decoupled limit
\begin{equation}
V_{\rm vdW}^{\rm dc} = -
\frac{4\varepsilon_0}{\ln^2(\pi\lambda/d)}\frac{T}{d\varepsilon_0}
\left(\frac{\lambda}{R}\right)^4.
\label{V4_vdW}
\end{equation}
The interlayer distance $d$ provides the large $k_z$ cutoff on the
thermal fluctuations of the individual lines.

In the continuous anisotropic case the single vortex elasticity is
still dominated by its electromagnetic contribution. At intermediate
distances $\lambda < R < d/\varepsilon$ we recover again the result
(\ref{V4_vdW}) as the Josephson coupling is not effective yet. At
large distances $\lambda, d/\varepsilon < R < \lambda_c$, however, the
interlayer coupling becomes important through cutting off the $k_z$
modes at $1/\varepsilon R$ and we find the result
\begin{equation}
V_{\rm vdW}^{\rm ret} = -
\frac{(171\pi/256)\varepsilon_0}{\ln^2(\pi\lambda_c/R)} \frac{T}
{\varepsilon\lambda\varepsilon_0}\left(\frac{\lambda}{R}\right)^5.
\label{V5_vdW}
\end{equation}
These results coincide with the ones found by Blatter and
Geshkenbein\cite{GB} up to the numerical prefactor as they missed a
factor $2$ in Eq.~(\ref{t^2}), as well as a term in the partial
Fourier transform which contributes to the result (\ref{V5_vdW}) to
order $O(\varepsilon^{-1})$. The same mistakes show up in their
results for the attraction of the vortex to the sample surface,
Eqs.~(19) and (20) of Ref.\ \onlinecite{GB}. We give the correct
expressions here:
\begin{equation}
V_{\rm vdW}^{\rm s,dc} = -\frac{\varepsilon_0/2}{\ln(\pi\lambda/d)}
\frac{T}{d\varepsilon_0}\left(\frac{\lambda}{R}\right)^2\nonumber
\end{equation}
for the decoupled case, and
\begin{equation}
V_{\rm vdW}^{\rm s} =
-\frac{(3/16)\varepsilon_0}{\ln(\pi\lambda/\varepsilon R)}
\frac{T}{\varepsilon\lambda\varepsilon_0}
\left(\frac{\lambda}{R}\right)^3
\nonumber
\end{equation}
in the continuous anisotropic situation.

\section{Casimir Force}

In this section, we derive the Casimir force between two parallel
cubes of vortex matter and then show, via the method of pairwise
summation, that the results coincide with those found by means of the
van der Waals approach.

Again, we begin with a simple analysis based on dimensional estimates,
instructing us what to expect. Consider two $d$-dimensional hypercubes
of size $L^d$ separated by a distance $R \ll L$, see Fig.~1. We wish
to determine the Casimir force $f$ per unit area acting between the
interfaces. Being due to the zero point fluctuations of the gauge
field, the only relevant dimensional quantities entering the
expression for the force are Planck's constant
$\hbar^{\scriptscriptstyle B}$, the velocity of light $c$, and the
distance $R$ between the slabs. The combination
$\hbar^{\scriptscriptstyle B}c/R^{d+1}$ then provides us with a
dimensionally correct expression for the force; in two dimensions we
then find $f \sim \hbar^{\scriptscriptstyle B}c/R^3$ for the retarded
Casimir effect. In the non-retarded case, $c \rightarrow \infty$ and
we have to replace the combination $c/R$ by the frequency scale
$\omega$ where the matter becomes transparent (for the vortex matter
in a layered superconductor this `frequency' is given by the layer
separation $d$, $\omega \sim 1/d$). Dimensional estimates then give us
for the force per unit length in two dimensions $f \sim
\hbar^{\scriptscriptstyle B} \omega /R^2$. In order to derive the
corresponding microscopic inter-particle potential from these
estimates, we apply the method of pairwise summation and obtain the
result $V_{\rm vdW} \propto 1/\varepsilon R^5$ in the retarded case
and $V_{\rm vdW} \propto 1/R^4$ in the non-retarded limit. These
simple estimates produce the correct power laws for the van der Waals
potential; in order to find the correct sign and the complete
prefactors, we have to go through the detailed analysis below.

\subsection{Formalism}

The Casimir energy is given through the difference in the sum over all
cavity modes minus the free field contribution\cite{Mostepanenko,PMG}.
While the calculation is rather straightforward for the simple case of
a metallic cavity, an appropriate formalism has to be set up to treat
more general configurations involving dielectric media.

We start from a quadratic Lagrangian density ${\cal L} = {\bf a} {\bf
  G}^{-1} {\bf a}$ in $d$ dimensions, with ${\bf a}({\bf x},\tau)$ a
vector boson field with $r$ components and ${\bf G}$ the Green
function matrix. The partition function ${\cal Z}$ is expressed
through the usual imaginary time path integral formalism
\begin{equation}
        {\cal Z}\!= \!\int\!{\cal D}[{\bf a}]\,
        \exp\left[-\frac{1}{\hbar} 
        \int {\rm d}^d x \int_0^{\hbar/T} 
        \!\!\!\!\!\!\!\!d\tau \, {\cal L}[{\bf a}]\right] 
        = [\det {\bf G}]^{-1/2},
\end{equation}
and the free energy $F = -T \ln{\cal Z}$ is given by
\begin{equation}
        F = \frac{T}{2} {\rm Tr}\ln {\bf G}^{-1}.
\end{equation}
We consider the classic parallel-plate geometry with isotropic media
separated by a gap of width $R$ along the $x_d$-direction. For each
polarization $\nu \in \{1, \dots , r\}$ we determine the eigenstates
obeying the boundary conditions at the $(d-1)$-dimensional
hypersurfaces placed at $0$ and $R$. The individual modes are
characterized by their polarization $\nu$, the (Matsubara) frequency
$\xi_s = 2\pi s T / \hbar$, $s \in {\mathbb{Z}}$, the transverse wave
vector ${\bf q} \in {\mathbb{R}}^{d-1}$, and the longitudinal wave
vector $k_n$, $n \in {\mathbb{}N}$. Given the set $(\nu, \xi_s, {\bf
  q})$, the discrete wave vectors $k_n$ are obtained as the solutions
of the boundary condition $D_{\{\alpha\}} (k)= 0$, where $\{\alpha\}$
is a short hand for the other indices $(\nu,\xi_s,{\bf q})$ (the
function $D_{\{\alpha\}} (k)$ derives from the determinant associated
with the set of boundary conditions; for the classic Casimir effect,
$D(k) = \sin(kR)$ with $R$ the distance between the metal plates).
Expressing the trace through all the indices we have
\begin{equation}
        F = \frac{T L^{d-1}}{2}\sum_{\nu=1}^r {\sum^{\infty}_{s=0}}
        \!\!\phantom{|}^\prime \!\int\! \frac{{\rm
        d}^{d-1}{q}}{(2\pi)^{d-1}} \!\!\!{\sum_{D_{\{\alpha\}} (k) =
        0}}\!\!\!  \ln G_{\{\alpha\}}^{-1}(k),
\label{F_ln}
\end{equation}
where the prime on the sum indicates that we count the $s=0$ term with
a weight $1/2$. In (\ref{F_ln}) we have made explicit the boundary
condition at the hypersurfaces separating the media from the gap.
\begin{figure}
\centerline{\epsfxsize= 5.5cm\epsfbox{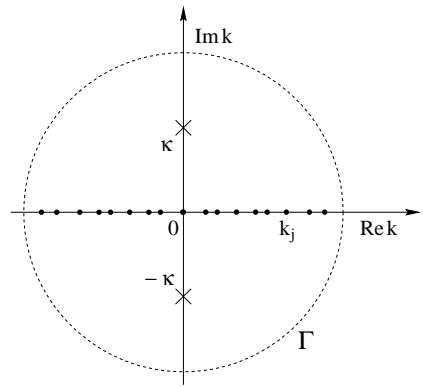}}
\vspace{0.5cm}
\caption{Contour and distribution of poles in the complex $k$-plane as
  they appear in the expression (\ref{C_Loop}). The poles produced by
  the boundary condition $D(k) = 0$ are marked by the solid dots on
  the real axis. The zeros of $g(k)$ are denoted by the crosses. All
  zeroes are distributed symmetrically with respect to the origin, a
  consequence of the symmetry $G(k) = G(-k)$.}
\label{fig:3}
\end{figure}
We make use of Cauchy's theorem to perform the sum over the
longitudinal momenta $k_n$: We first rewrite the sum in the form
\[
        {\sum_{D_{\{\alpha\}}(k)=0}} \ln[G_{\{\alpha\}}^{-1} + y] 
       = \int {\rm d}y \sum_{D_{\{\alpha\}}(k)=0} 
         \frac{1}{G_{\{\alpha\}}^{-1} + y}.
\]
Next, we let the boundary condition $D_{\{\alpha\}} (k)= 0$ produce
the desired sum in a Cauchy loop integral in the complex $k$ plane,
see Fig.\ 3. With $g(k) = (G_{\{\alpha\}}^{-1}(k) + y)$ we have
\begin{eqnarray}
&&\frac{1}{2\pi i} \oint_\Gamma \frac{\partial_k D(k)}{D(k)}
\frac{1}{g(k)} \label{C_Loop} \\
&&= \sum_{D(k)=0} \frac{1}{g(k)} 
  + \sum_{g(k)=0} \frac{\partial_k D(k)}{D(k)} 
                  \frac{1}{\partial_k g(k)},
\nonumber
\end{eqnarray}
where it is understood that both $D$ and $g$ further depend on the
other indices $\{\alpha\}$ as well as on the parameter $y$ in the case
of $g$. We define the function $w(\{\alpha\}; y)$ as the zeroes of the
expression $g_{\{\alpha\}}(k) = G_{\{\alpha\}}^{-1}[w(\{\alpha\};y)] 
+ y =0$. Taking the derivative of the last equation with respect 
to $y$, $\partial_y g = \partial_y G^{-1} + 1 = 
\partial_k G^{-1} \partial_y w + 1$, and using the result
in (\ref{C_Loop}), we obtain the desired result
\begin{equation}
        \sum_{D_{\{\alpha\}}(k) =0} \frac{1}{G_{\{\alpha\}}^{-1} + y}
        = 2\frac{\partial}{\partial y} \ln
        D_{\{\alpha\}}[w(\{\alpha\};y)],
\label{sum_k}
\end{equation}
where we have used that the Green function exhibits the proper
asymptotics $G(k) \sim O(k^2)$ and the symmetry $G(k) = G(-k)$.
Inserting the sum on $k$, Eq.\ (\ref{sum_k}) with $y = 0$, back into
the free energy expression (\ref{F_ln}) we obtain
\begin{equation}
        {\cal F} = T \sum_{\{\alpha\}} \ln
        D_{\{\alpha\}}[w(\{\alpha\})],
\label{F_lnD}
\end{equation}
where 
\[
        \sum_{\{\alpha\}} \, [\dots] \equiv L^{d-1}\sum_{\nu =1}^r
        {\sum^{\infty}_{s=0}} \!\!\phantom{|}^\prime \int \frac{{\rm
        d}^{d-1}q}{(2\pi)^{d-1}} \, [\dots]
\]
denotes the sum/integral over all modes $\{\alpha\} = (\nu,\xi_s,{\bf
q})$.

The above formal manipulations require a regularization such that
$D[\xi_s \rightarrow \infty] \rightarrow 1$. We carry out the
appropriate renormalization by the subtraction of the free energy of
the free field without boundary conditions, corresponding to replacing
the partition function through the ratio ${\cal Z} \rightarrow {\cal
  Z}/{\cal Z}_0$, with ${\cal Z}_0$ the partition function of the free
field. For the latter, we define the `boundary condition'
$D_0$ in such a way as to produce the fraction $R/L$ of the free field
energy in the volume $L^d$ via the expression (\ref{F_lnD});
for the classic Casimir problem, $D_0(k\pm i\delta) \propto 
\lim_{L \rightarrow \infty} \{\sin[(k\pm i\delta)L]\}^{R/L}$.
With $D^{\rm ren} = D/D_0$ we finally obtain the regularized
free energy of the Casimir problem\cite{Mostepanenko}
\begin{equation}
        {\cal F}^{\rm ren} = T \sum_{\{\alpha\}} 
         \ln D_{\{\alpha\}}^{\rm ren}[w(\{\alpha\})].
\label{fren}
\end{equation}

We illustrate the use of this formalism with a brief derivation of the
classic Casimir result, the attraction between two parallel metallic
plates of size $L\times L$ and a distance $R$ apart. The dispersion
relation for the free electromagnetic field is $G^{-1} (\xi_s, {\bf
  q}, k) = \xi_s^2/c^2 + q^2 + k^2$, such that the function $w|_{y=0}$
takes the form $w(\xi_s, {\bf q}) = i \sqrt{\xi_s^2/c^2 + q^2}$. The
boundary condition requiring the fields to vanish at $0$ and $R$ can
be cast into the form $D(k) = \sin(kR) = 0$, producing the modes with
a longitudinal $k$ vector $k_n = 2 \pi n/R$. The `boundary condition'
describing the free field takes the form $D_0(k \pm i\delta)
= (\mp 1/2i) \exp(\mp i k R)$, such that $D^{\rm ren}(k \pm i\delta) =
1 - \exp(\mp 2 i k R)$, where $\delta \rightarrow 0^+$. Combining the
results for $D^{\rm ren}$ and $w$, we have to carry out the mode
summation over the logarithm of
\begin{equation}
D^{\rm ren}[w(\xi_s,{\bf q})] = 
\left[1-\exp\left(-2R \sqrt{\frac{\xi_s^2}{c^2}+q^2}\right)\right]^2,
\end{equation}
where taking the square accounts for the two polarization modes of the
electromagnetic field (the single polarization mode at $k=0$ has been
properly taken into account here). The force density $f$ between the
plates is given through the derivative of the energy (\ref{fren}) with
respect to $R$,
\begin{eqnarray}
f &=& - \frac{1}{L^2} \frac{\partial {\cal F}^{\rm ren}}{\partial R} =
- \frac{\hbar}{\pi R}\int_0^\infty 
{\rm d}q q \int_0^\infty \frac{{\rm d} \xi}{2 \pi} \nonumber \\ 
&& \qquad \times
\frac{2R\sqrt{\xi^2/c^2+q^2}}{\exp(2R\sqrt{\xi^2/c^2+q^2})-1},
\label{force_density}
\end{eqnarray}
where we have replaced the Matsubara sum through an integral over
complex frequencies $\xi$. The remaining integrals are easily
evaluated by changing to polar coordinates $q = \rho \cos \phi$,
$\xi/c = \rho \sin \phi$ and we obtain the classic result due to
Casimir\cite{Casim}
\begin{equation}
f = - \frac{\pi^2}{240}\frac{\hbar c}{R^4}.
\end{equation}

\subsection{Application to 2D bosons}

In the following, we apply the above formalism to the 2D charged
bosons described by the action ${\cal S} = {\cal S}_0 + {\cal S}_{\rm
  int}$ as given in Eqs. (\ref{plm}) and (\ref{edr}). Following the
scheme outlined in Fig.\ 2, we first have to integrate over the
currents ${\bf J}\sim \partial_\tau {\bf R}$. We split the boson
positions ${\bf R}_\mu(\tau)$ into the mean field part ${\bf
  R}^0_\mu(\tau)$ and a fluctuating part ${\bf u}_\mu(\tau)$, ${\bf
  R}_\mu(\tau) = {\bf R}_\mu^0 (\tau) + {\bf u}_\mu(\tau)$. In
carrying out the integration over the boson positions ${\bf
  R}_\mu(\tau)$ we can ignore the fluctuating part ${\bf u}_\mu(\tau)$
in the gauge field ${\bf a}$ as the latter is smooth on the scale of
the amplitude ${\bf u}_\mu(\tau)$,
\begin{eqnarray}
&& \int {\cal D}[{\bf u_\mu}] 
\exp\Bigg\{-\frac{1}{\hbar^{\scriptscriptstyle B}}
\int{\rm d}\tau\sum_\mu
\Big[\frac{m}{2}(\partial_\tau {\bf u}_\mu)^2 
\nonumber \\ &&
\qquad\qquad\qquad\qquad\qquad
+ i(\partial_\tau {\bf u}_\mu,1)\cdot 
{\bf a}({\bf R}_\mu^0,\tau)\Big]\Bigg\}
\nonumber \\
&& \propto \exp\left[-\frac{1}{\hbar^{\scriptscriptstyle B}}
\!\! \int \! {\rm d}^2 R{\rm d}\tau 
\left(\frac{1}{2} \Pi({\bf R}) a^2_{xy}({\bf R}) + 
i \rho({\bf R}) a_\tau({\bf R}) \right)\right],
\nonumber
\end{eqnarray}
where
\[
\Pi({\bf R}) = \frac{\rho({\bf R})}{m} \quad {\rm and} \quad 
\rho({\bf R}) = \sum_\mu \delta^{(2)} ({\bf R} - {\bf R}^0_\mu)
\]
denote the polarizability and the density of the 2D bosons,
respectively.  Second, we have to integrate over the true gauge field
${\bf A}$ (we have fixed the gauge to $\nabla \cdot {\bf a} = 0$). The
resulting term $(1/2 g^2 \lambda^2) a^2$ in the action then renders
the fake gauge field ${\bf a}$ massive (physically, this mass term
expresses the finite range $\lambda$ of the interaction between the
charged bosons/vortices). Third, we introduce the free Green function
matrix ${\bf G}_0$ for the gauge field, in Fourier representation,
\begin{eqnarray}
{\bf G}_0^{-1} &=& 
\left[\begin{array}{cc} {\bf g}_{xy}^{-1} & 0 \\ 0 &
g_\tau^{-1} \end{array} \right], \nonumber \\ 
\noalign{\vskip 5 pt}
{\rm with} \quad ({\bf g}_{xy})_{\alpha \beta}^{-1} 
&=& k^2 \delta_{\alpha\beta} + (c^2-1) K^2
P_{\alpha \beta}({\bf K}), \nonumber \\
g_\tau^{-1} &=& k^2, \nonumber
\end{eqnarray}
and the transverse projector $P_{\alpha\beta}=\delta_{\alpha\beta}-
K_\alpha K_\beta/K^2$. Combining the results of the above three steps,
we arrive at the desired effective action for the fake gauge field
${\bf a}$,
\begin{eqnarray}
{\cal S}_{\rm eff}[{\bf a}] &=& \int\!\!{\rm d}^2 R {\rm d}\tau
\left[
{\bf a}_{xy} \frac{1}{2g^2}\left({\bf g}_{xy}^{-1} + 
\frac{1}{\lambda^2} + g^2\Pi \right) {\bf a}_{xy} 
\right. 
\nonumber \\ &&
\left. 
+a_\tau \frac{1}{2g^2}\left(g_\tau^{-1} + 
\frac{1}{\lambda^2}\right)a_\tau + i\rho a_\tau
\right].
\label{thing}
\end{eqnarray}
The Euler-Lagrange equations of this action determine the (imaginary
time) field equations for the transverse and longitudinal parts of
${\bf a}$ (we remind that $\lambda_c = c \lambda =
\lambda/\varepsilon$),
\begin{eqnarray}
&&\hspace{-0.1cm}\left(\frac{1}{c^2} \partial_\tau^2
+{{\nabla}_{xy}}^2 
-\frac{g^2\rho}{m^{\scriptscriptstyle B}c^2}-
\frac{1}{\lambda_c^2} \right){\bf a}_{xy}
\nonumber \\
&&\hspace{2.0cm}= \left(1 - \frac{1}{c^2}\right){{\nabla}_{xy}}
({{\nabla}_{xy}}\cdot {\bf a}_{xy}) = 0
\label{trans_eom}
\end{eqnarray}
and
\begin{equation}
\left(\nabla_{xy}^2 - \frac{1}{\lambda^2}\right)a_\tau = i\rho g^2,
\label{a0_eqn}
\end{equation}
with $\nabla_{xy} = (\partial_x, \partial_y)$. In the last equation of
(\ref{trans_eom}) we have made use of the gauge condition
$\nabla\cdot{\bf a}={\nabla}_{xy}\cdot{\bf a}_{xy} +\partial_\tau
a_\tau=0$ and have ignored the time dependence in the longitudinal
field component $a_\tau$, see below.

The longitudinal field $a_\tau$ is generated by the source $g^2 \rho
/4 \pi = \varepsilon_0 \rho$ and has no dynamics on its own.  Placing
two bosons at a distance $R$, a simple integration gives the
vortex-vortex interaction (c.f., Eq.~(\ref{rep_eq}))
\begin{equation}
V_{\rm rep}(R) = 2\varepsilon_0 K_0(R/\lambda).
\label{vv_int}
\end{equation}
Here, we ignore the fluctuations of the vortex position in the
calculation of $a_\tau$ as they merely produce a small renormalization
of the prefactor in the repulsive potential.  Similarly, we can
neglect such corrections in the calculation of the transverse modes
below and set the r.~h.~s.~of (\ref{trans_eom}) equal to zero.

The dynamical transverse field ${\bf a}_{xy}$ generates the Casimir
force. Here, we have in mind a geometry as sketched in Fig.\ 1, with
two parallel cubes of vortex matter of density $\rho$, separated by a
vortex free region of thickness $R$. We choose the $x$-axis to lie
perpendicular to the `plates', the $z$-axis is directed along the
vortices. Within the boson language we deal with two parallel planes
of 2D bosons lying in the $xy$-plane, the direction along the vortices
now transforming to the imaginary time coordinate (below, we consider
the limit $T^{\scriptscriptstyle B} \rightarrow 0$, implying vortex
lines of infinite length, $L_z \rightarrow \infty$). Note that in the
present 2D case we have to consider only one polarization mode for the
transverse gauge field ${\bf a}_{xy}$, a consequence of the gauge
condition $\nabla \cdot {\bf a} = 0$.

Going over to Fourier space we can cast (\ref{trans_eom}) into the
form (we remind the reader that $c = 1/\varepsilon$; $q$ and $k$ are
the wave vectors along and perpendicular to the `plates')
\begin{equation}
\left[\frac{\xi^2}{c^2}\left(1+\frac{g^2\rho}{m^{\scriptscriptstyle
B}\xi^2} \right) + q^2 + k^2 + \frac{1}{\lambda_c^2} \right] {\bf
a}_{xy} = 0,
\label{trans_eom_k}
\end{equation}
from which we obtain the Green function $G_0^{-1}(\xi, q, k) =
\epsilon_{\scriptscriptstyle V} (\xi) \xi^2/c^2 + q^2 + k^2 +
\lambda_c^{-2}$ and the function $w$,
\begin{equation}
w(\xi, q) = i \sqrt{\frac{\epsilon_{\scriptscriptstyle V}
(\xi)}{c^2}\xi^2 + q^2 + \frac{1}{\lambda_c^2}}.
\label{fvdW}
\end{equation}
Here, we have defined the `dielectric' constant
\begin{equation}
\epsilon_{\scriptscriptstyle V} (\xi) = 
1 + \frac{g^2\rho}{m^{\scriptscriptstyle B}
\xi^2} = 1 + \frac{8\pi\lambda^2\rho}{\ln(1+\lambda^2\xi^2)},
\label{epsilon_v}
\end{equation}
where we go over to vortex parameters in the last equation
($m^{\scriptscriptstyle B} \rightarrow \varepsilon_l$) and make use of
the expression (\ref{elasticity}) for the dispersive vortex elasticity
$\varepsilon_l$.

Next, we have to formulate the boundary conditions in terms of the
zeroes of the function $D_{\{\alpha\}}(k) = 0$. The translational
invariance in the $y$-direction allows for a plane-wave Ansatz ${\bf
  a}_{xy}(y) \propto e^{i q y}$, while the sequence of `dielectric'
and `vacuum' regions along the $x$-axis, requires to match the plane
waves ${\bf a}_{xy}(x) \propto e^{i k(x) x}$, with (see
(\ref{trans_eom_k}))
\begin{eqnarray}
k_l &=& k(x<0) = \sqrt{k^2
- \frac{(\epsilon_{\scriptscriptstyle V}(\xi)-1)}{c^2}\xi^2}, 
\nonumber \\
k_0 &=& k(0<x<R) = k, \nonumber \\
k_r &=& k(R<x) = \sqrt{k^2 
- \frac{(\epsilon_{\scriptscriptstyle V}(\xi)-1)}{c^2}\xi^2}.
\label{ki}
\end{eqnarray}
The gauge condition $\nabla_{xy} \cdot {\bf a}_{xy}$ is satisfied
through the An\-satz
\begin{eqnarray}
{\bf a}_{xy} &=& 
\alpha(x) \left(\begin{array}{c} q \\-k(x) \end{array}\right) 
e^{i(k(x)x+qy)} 
\nonumber \\
&& \qquad + \beta(x)  \left(\begin{array}{c} q \\ k(x) 
\end{array} 
\right)e^{-i(k(x)x+qy)}, 
\nonumber
\end{eqnarray}
with piecewise constant amplitudes $\alpha(x)$ and $\beta(x)$.  The
six coefficients are determined by the boundary conditions at $x=0$
and $x=R$, requiring the parallel electric field ${\bf e}_\parallel$
and the magnetic field $b$ to be continuous, see Appendix B. Two
further conditions force the fields to vanish at $x = \pm L$, $L
\rightarrow \infty$. Requiring the determinant of the resulting $6
\times 6$ matrix problem to vanish, we find
\begin{eqnarray}
   D[\xi,q,k] &=& e^{-ik_l L} e^{-ik_0 R} e^{-ik_r L} \label{DvdW} \\
&&\left[1 - \frac{
  (1+ (\lambda\xi)^{-2})k_l-(\epsilon_{\scriptscriptstyle V}+
   (\lambda\xi)^{-2})k 
               }{
  (1+ (\lambda\xi)^{-2})k_l+(\epsilon_{\scriptscriptstyle V}+ 
   (\lambda\xi)^{-2})k
               }\right. 
  \nonumber \\ && \times  \left.
          \frac{
  (1+ (\lambda\xi)^{-2})k_r-(\epsilon_{\scriptscriptstyle V}+
  (\lambda\xi)^{-2})k
               }{
  (1+ (\lambda\xi)^{-2})k_r+(\epsilon_{\scriptscriptstyle V}+
  (\lambda\xi)^{-2})k
                } e^{2 i k R}\right].
\nonumber
\end{eqnarray}
For the free field, we find 
\begin{equation}
D_0(k+i\delta) =  e^{-ik_l L} e^{-ik_0 R} e^{-ik_r L},
\label{DvdW_0}
\end{equation}
and combining the above results for $w$, $D$, and $D_0$,
Eqs.~(\ref{fvdW}), (\ref{DvdW}), and (\ref{DvdW_0}), we can construct
the function $D^{\rm ren}[w(\xi,q)]$ and obtain the formal expression
for the free energy (\ref{fren}) of the vortex Casimir problem.  We
note that the cut introduced through the definitions of $k_l$ and
$k_r$ in Eq.\ (\ref{ki}) disappears from the boundary condition $D$
and hence does not contribute to the loop integral (\ref{C_Loop}). As
a result, the free energy expression (\ref{fren}) for the Casimir free
energy remains valid.

\subsection{Casimir Interaction}

In the present two dimensional situation the expression for the
Casimir energy Eq.~(\ref{fren}) per length reads
\begin{equation}
\frac{{\cal F}^{\rm ren}(R)}{L} = \hbar^{\scriptscriptstyle B}
\int_0^\infty \frac{{\rm d} q} {2\pi}
\int_0^\infty \frac{{\rm d}\xi}{\pi}\ln D^{\rm ren}[w(\xi,q)],
\end{equation}
and produces the Casimir force per length $L$ 
\begin{eqnarray}
f &=& -\frac{1}{L}\frac{\partial {\cal F}^{\rm ren}(R)}{\partial R}
\nonumber \\
&=&-\frac{\hbar^{\scriptscriptstyle B}}{i\pi^2}\int_0^\infty{\rm d}q 
   \int_0^\infty {\rm d}\xi \, w(\xi,q) \label{force_gen_0} \\
&&    \qquad\qquad\times \displaystyle{
   \Bigg[\frac{(1+(\lambda\xi)^{-2})w_l+
        (\epsilon_{\scriptscriptstyle V}+(\lambda\xi)^{-2})w}
              {(1+(\lambda\xi)^{-2})w_l-
        (\epsilon_{\scriptscriptstyle V}+(\lambda\xi)^{-2})w}
                     } \nonumber \\
&&      \displaystyle{\times
         \frac{(1+(\lambda\xi)^{-2})w_r+
        (\epsilon_{\scriptscriptstyle V}+(\lambda\xi)^{-2})w}
              {(1+(\lambda\xi)^{-2})w_r-
        (\epsilon_{\scriptscriptstyle V}+(\lambda\xi)^{-2})w}
   e^{-2iwR} - 1 \Bigg]^{-1}
                   }. \nonumber
\end{eqnarray}
We introduce the new variables
\begin{eqnarray}
p &=& \sqrt{1 + c^2 q^2/\xi^2}, \nonumber \\ 
s &=&\sqrt{p^2+(\lambda\xi)^{-2}}, \nonumber \\
s_\epsilon &=&\sqrt{(\epsilon_{\scriptscriptstyle V}-1)+p^2 + 
                                     (\lambda\xi)^{-2}}, 
\nonumber
\end{eqnarray}
and arrive at the simplified expression 
\begin{eqnarray}
&&   f =\displaystyle{
-\frac{\hbar^{\scriptscriptstyle B}}{\pi^{2}}\int_1^{\infty}{\rm d}p 
\int_{0}^{\infty}{\rm d}\xi \frac{s p \xi^2}{c^2 \sqrt{p^2-1}} 
                     } \label{force_gen} \\ 
&&   \displaystyle{
   \left[\left(\frac{(1+(\lambda\xi)^{-2})s_\epsilon+
     (\epsilon_{\scriptscriptstyle V}+(\lambda\xi)^{-2})s} 
           {(1+(\lambda\xi)^{-2})s_\epsilon-
     (\epsilon_{\scriptscriptstyle V}+(\lambda\xi)^{-2})s}\right)^2 
   e^{2 \xi s R/c} - 1 \right]^{-1}
                   }. \nonumber
\end{eqnarray}
Eq.~(\ref{force_gen}) gives the two dimensional version for massive
photons of the Casimir force between dielectric media, first derived
by Lifshitz\cite{Lif} for the conventional 3D situation.

In the following we will be interested in the situation where the
`dielectric' media are dilute (dilute vortex matter at small magnetic
induction $B \ll \Phi_0/\lambda^2$). In this case, we expand
(\ref{force_gen}) in the small correction
$\epsilon_{\scriptscriptstyle V}-1$ giving the deviation of the
dielectric constant from its vacuum value. Furthermore, if the
integral is dominated by large frequencies $\xi$ we can ignore the
mass term everywhere. Under these conditions, we approximate $s
\approx p$ and $s_\varepsilon \approx (\epsilon_{\scriptscriptstyle V}
- 1) (1/2p - p) + \epsilon_{\scriptscriptstyle V} p$, such that
$s_\varepsilon - \epsilon_{\scriptscriptstyle V} p \approx
(\epsilon_{\scriptscriptstyle V} - 1) (1/2p - p)$, while
$s_\varepsilon + \epsilon_{\scriptscriptstyle V} p \approx 2p$, and
obtain the simplified force expression (the weak logarithmic disperion
in $\epsilon_{\scriptscriptstyle V}$ is approximated through an
appropriate constant)
\begin{equation}
f =-\frac{\hbar^{\scriptscriptstyle B}
(\epsilon_{\scriptscriptstyle V}-1)^2}{16\pi^2}\!\!\int_1^\infty 
\!\!\! {\rm d}p \int_0^{\infty}\!\!\!\!{\rm d}\xi 
\frac{\xi^2 (2p^2-1)^2}{c^2 p^2 \sqrt{p^2-1}} e^{2 \xi p R/c}
\label{force_dil} 
\end{equation}

In (\ref{force_gen}), (\ref{force_dil}) we have to account for three
relevant frequency- or length scales. The first is introduced 
by the discrete nature of our superconductor: the layered structure
limits the $\xi$-integration through the frequency cutoff 
$\xi_d = \pi / d$ (this corresponds to the medium becoming
transparent at high frequencies $\xi > \xi_d$). The second
frequency scale is introduced through the dispersion in
the dielectric constant: at low frequencies $\xi < \xi_\lambda 
= 1/\lambda$, the dielectric constant takes the form
$\epsilon_{\scriptscriptstyle V} (\xi \rightarrow 0) \rightarrow
8\pi\rho/\xi^2$ and the dielectric response changes
over to a mass renormalization. The third
frequency scale $\xi_R = 1/R\varepsilon$ appears through the
exponential $\exp(2 \xi s R/c) > \exp(2\xi R \varepsilon)$ in the
integrand of (\ref{force_gen}); as $\xi$ goes beyond $\xi_R$ the
exponent cuts off the integrand in (\ref{force_gen}).

Thus, depending on the frequency $\xi$, the media are either
transparent, give a dielectric response, or produce a mass
renormalization.  And depending on the position of $\xi_R$ with
respect to these response regimes, we will find a different behavior
of the Casimir force. Below, we will analyze the various regimes and
use the pairwise summation technique to show that the results are in
agreement with those obtained for the van der Waals interaction
potential in Sec.\ IV above.

\subsubsection{Intermediate Distances $\lambda < R < d/\varepsilon$}

At intermediate distances $\lambda < R < d/\varepsilon$, we have
$\xi_d < \xi_R$ such that the cutoff on the $\xi$ integral is given by
$\xi_d$. As the integral is dominated by large values of $\xi$ we can
ignore the mass terms $1/\lambda^2\xi^2$ everywhere. The exponential
$\exp(2 \xi s R/c)$ restricts the $p$-integral to values $p <
d/\varepsilon R$, admitting large values of $p$ such that we can drop
the corrections to $p^2$ in (\ref{force_dil}), e.g., $p^2 - 1 \approx
p^2$. Transforming $p$ to the new variable $\gamma = 2 R \xi \, p/c$,
we find the Casimir force on intermediate scales
\begin{eqnarray}
f &=& - \frac{\hbar^{\scriptscriptstyle B}
(\epsilon_{\scriptscriptstyle V}-1)^2}{16\pi^{2}R^2} 
\int_{0}^{\xi_d}  {\rm d}\xi \int_{0}^{\infty} {\rm d}\gamma \gamma 
e^{-\gamma} \nonumber\\
 &=& - \frac{(\epsilon_{\scriptscriptstyle V}-1)^2}{16\pi^{2}} 
\frac{\hbar^{\scriptscriptstyle B}\xi_{d}} {R^{2}}.
\label{shorcasimir}
\end{eqnarray}
Using the result (\ref{F_of_R}) of the pairwise summation and
inserting the expression (\ref{epsilon_v}) for the dielectric
constant, we obtain the van der Waals interaction describing the
decoupled limit,
\begin{equation}
V_{\rm vdW}^{\rm dc} = -
\frac{4\varepsilon_0}{\ln^2(\pi\lambda/d)}\frac{T}{d\varepsilon_0}
\left(\frac{\lambda}{R}\right)^4,
\label{shorvdW}
\end{equation}
in agreement with the result (\ref{V4_vdW}) (note that the density
$\rho$ cancels in the final result for the van der Waals interaction).

\subsubsection{Large Distances $\lambda, d/\varepsilon < R < 
\lambda/\varepsilon$}

At large distances, the $\xi$-integral is cutoff on the scale $\xi_R$,
i.e., the Casimir force is limited by the distance $R$ through the
exponential $\exp(2 \xi s R/c)$ rather than by the transparency of the
material.  We still can ignore the mass terms in (\ref{force_gen}) as
we assume that $\xi_\lambda < \xi_R$, providing us with a large regime
for the $\xi$-integration where $\xi > 1/\lambda$.  Transforming the
energy variable $\xi$ to $\gamma = (2R\xi p)/c$, we obtain the
following expression for the Casimir force in the low density limit,
\begin{eqnarray}
f &=& - \frac{\hbar^{\scriptscriptstyle
B}c(\epsilon_{\scriptscriptstyle V}-1)^2} {8\pi^2 R^3}
\int_{0}^{\infty} {\rm d}\gamma \gamma^2 e^{-\gamma} \int_1^\infty
{\rm d}p \frac{(2p^2-1)^2}{16p^5\sqrt{p^2-1}} \nonumber \\ &=& -
\frac{19(\epsilon_{\scriptscriptstyle V}-1)^2}{1024 \pi}
\frac{\hbar^{\scriptscriptstyle B} c}{R^3}.
\label{longforce}
\end{eqnarray}
Following the usual scheme, this result produces the retarded van der
Waals interaction
\begin{equation}
V_{\rm vdW}^{\rm ret} = -\frac{(171\pi/256)\varepsilon_0}
{\ln^2(\pi\lambda/\varepsilon R)}\frac{T}{\lambda\varepsilon
\varepsilon_0}\left(\frac{\lambda}{R}\right)^5,
\label{long_vdW}
\end{equation}
in agreement with Eq.~(\ref{V5_vdW}).

\subsubsection{Very Large Distances $\lambda/\varepsilon < R$}

The situation at very large distances $R> \lambda / \varepsilon =
\lambda_c$ is most conveniently analyzed in the original formulation
(\ref{force_gen_0}). The mode summation is limited by the exponential
$\exp(2iwR)$, implying the following restrictions on the integration
variables $\xi$ and $q$: $\xi/c < 1/\sqrt{R\lambda_c} < 1/\lambda_c$
and $q < 1/\sqrt{R\lambda_c} < 1/ \lambda_c$.  The integral
$\int_0^\infty {\rm d} q \int_0^\infty {\rm d} \xi \, [\dots]$ then is
given by the area $c/R\lambda_c$ times the $q,\xi \rightarrow 0$ limit
of the integrand. In the limit $\xi \rightarrow 0$ the mass terms are
relevant and the dispersion in the dielectric constant
(\ref{epsilon_v}) produces the additional mass renormalization
\begin{equation}
\frac{1}{\lambda_c} \rightarrow \sqrt{\frac{1}{\lambda_c^2} + 
\frac{8\pi}{c^2}\rho} \equiv M.
\end{equation}
The Casimir force then decays exponentially following the law
\begin{equation}
f=-\frac{1}{2\pi}\frac{\hbar^{\scriptscriptstyle B}c}{\lambda_c^2 R}
\frac{(1- \lambda_cM)^2}{(1+\lambda_cM)^2} \, e^{-2R/\lambda_c}.
\end{equation}
For small densities, we can expand $\lambda_cM \approx 1 +
4\pi\lambda^2\rho$ and obtain for the screened van der Waals
interaction in the regime $\lambda/\varepsilon < R$ the expression
(note that the pairwise summation formula (\ref{F_of_R}) has to be
modified for the exponential factor in the interaction $V{\rm vdW}$)
\begin{equation}
V_{\rm vdW}^{\rm sc} 
= -4 \sqrt{\pi} \varepsilon_0 \frac{T\varepsilon^4}
{\varepsilon_0 \lambda} \left(\frac{\lambda_c}{R}\right)^{3/2}
\exp(-2R/\lambda_c).
\end{equation}

\section{Discussion}

We have derived the Casimir force between two bodies of vortex matter
and have inferred from the result the strength of the van der Waals
interaction between vortex lines via the method of pairwise summation.
While the calculation of the Casimir force is carried out in the boson
formulation, the van der Waals attraction is determined in the vortex
picture. The agreement between the results is once more an
illustration of the equivalence of the two formalisms 
\cite{Nelson,NelsonVinokur}.

The physical implications of these results lead to interesting
modifications of the $B$-$T$ phase diagram of layered/anisotropic type
II superconductors\cite{GB}: The attraction between the lines produces a
generic vortex solid of density $\rho \sim 10^{-2}/\lambda^2$ at low
temperatures. An interesting problem appearing in this context
is the accurate determination of the entropic repulsion between 
the vortex lines, the latter giving an important contribution 
to the Gibbs free energy. Various approaches to this problem 
have been discussed by Blatter and Geshkenbein\cite{BGvdWpC},
by Volmer, Mukherji, and Nattermann\cite{VMN}, and by Volmer and
Schwartz\cite{VS}. The transition from the Meissner state to
the van der Waals vortex solid takes place via a sharp
first-order transition at the lower critical field $H_{c_1}$.
The concurrent phase separation then can be observed in a decoration
experiment, though very clean samples are required. Furthermore,
the phase separation leads to the concentration of the vortex lines 
into dense regimes separated by vortex free regions. The interaction
between the vortex domains then is given by the Casimir force
calculated here, thus providing a natural realization of the
physics described in this paper.

A further remark is in place concerning the physical character
of the van der Waals interaction discussed here. The usual
van der Waals interaction between neutral atoms arises from
dipolar fluctuations in the charge distribution of the atoms,
resulting in a force which is mediated through the scalar
(longitudinal) potential, at least within the non-retarded regime 
at small distances. Comparing to the van der Waals force between 
the 2D charged bosons/vortex lines discussed here, we note that
the elementary objects do not have an internal structure
producing a fluctuating dipole. The van der Waals interaction
then arises from {\it current} fluctuations and thus involves
the (transverse) vector field. In principle, such a `transverse' 
van der Waals force is also present in an electron gas: while the
longitudinal repulsive interaction is screened on the Thomas-Fermi
length, the transverse attractive interaction of the van der Waals
type due to fluctuating currents survives at long distances (and
in principle induces a superconducting instability). However,
the transverse van der Waals force in an electronic system
involves the small parameter $v_{\scriptscriptstyle F}/c$
($v_{\scriptscriptstyle F}$ is the Fermi velocity), rendering
the effect small. In the 2D boson system discussed here, the
light velocity is given by the anisotropy of the superconductor,
$c = 1/\varepsilon$, and thus is a tunable parameter.

We thank Vadim Geshkenbein, A.~van Otterlo, and Christoph Bruder 
for fruitful discussions.

\begin{appendix}

\section{Partial Fourier Transforms}

In this section we sketch, by way of example, the calculation of
$V_{xx}^{\rm int}({\bf R},k_z)$. We start from Eq.~(\ref{Vab}) and set
$\alpha = \beta = x$ (here, we ignore the exponential cutoff function
due to the vortex core; the cases $\alpha = \beta = y$ and $\alpha =
x$, $\beta = y$ follow trivially)
\[
V_{xx}^{\rm int}({\bf K},k_z) = \frac{1}{1+ \lambda^2 k^2}\left(1 -
\frac{(\lambda_c^2 -\lambda^2)k_y^2}{1 + \lambda_c^2 K^2 +
\lambda^2k_z^2}\right).
\]
Performing a contour-integral over $k_x$, we obtain
\begin{eqnarray}
V_{xx}^{\rm int}({\bf R},k_z) &:=& I_1 + I_2 \nonumber \\ &&
\hspace{-2.0cm}=\frac{1}{4\pi \lambda a^2}\int {\rm d}k_y Y(\lambda,
k_y)e^{-(R_x / \lambda)Y(\lambda, k_y)} e^{ik_y R_y} \nonumber \\ &&
\hspace{-1.8cm}-\frac{\lambda_c}{4\pi a^2} \int {\rm d}k_y
\frac{k_y^2}{Y(\lambda_c, k_y)}e^{-(R_x /\lambda_c)Y(\lambda_c,k_y)}
e^{i k_yR_y}, \nonumber
\end{eqnarray}
with $Y(v,w) = \sqrt{a^2+ v^2w^2}$ and $a^2 = 1 + \lambda^2 k_z^2$.
Carrying out the integral\cite{GR} over $k_y$ in $I_1$ we find
\[
I_1 = \frac{1}{2\pi a^2}\frac{\partial^2}{\partial
R_x^2}K_0\left(\frac{aR}{\lambda}\right).
\]
The second integral $I_2$ can be calculated in the same manner,
\[
I_2 = \frac{1}{2\pi a^2}\frac{\partial^2}{\partial
R_y^2}K_0\left(\frac{aR}{\lambda_c}\right).
\]
Here, $K_0$ denotes the $0$-th order modified Bessel functions of the
second kind and $R^2 = R_x^2 + R_y^2$. Collecting results and carrying
out the derivatives, we obtain the desired result. Here, we quote the
final expressions for the case ${\bf R} = (R,0)$:
\begin{eqnarray}
  V_{xx}^{\rm int}(R,k_z) &=& 
  \frac{1}{2 \pi a^2}
\left\{\frac{a^2}{\lambda^2}\left[K_0\left(\frac{aR}{\lambda}\right)
              +\frac{\lambda}{aR}K_1\left(\frac{aR}{\lambda}\right)
                             \right] 
  \right. 
\nonumber  \\ && 
  \left. \qquad\qquad\qquad 
-\frac{a}{\lambda_c R} K_1\left(\frac{aR}{\lambda_c}\right)
  \right\},
\label{app_a_x} \\
  V_{yy}^{\rm int}(R,k_z) &=&
  \frac{1}{2 \pi a^2}
  \left
\{\frac{a^2}{\lambda_c^2}\left[K_0\left(\frac{aR}{\lambda_c}\right)
          +\frac{\lambda_c}{aR}K_1\left(\frac{aR}{\lambda_c}\right)
                             \right]
  \right.
\nonumber  \\ &&
  \left. \qquad\qquad\qquad
-\frac{a}{\lambda R} K_1\left(\frac{aR}{\lambda}\right)
  \right\}.
\label{app_a_y}
\end{eqnarray}
The third component $V_{xy}^{\rm int}(R,k_z)$ vanishes for $R_y =0$.
The important terms in (\ref{app_a_x}) and (\ref{app_a_y}) are those
involving the large screening length $\lambda_c$ in the argument of
the Bessel functions. The first term $\propto K_0$ in (\ref{app_a_y})
has been missed in Ref.\ \onlinecite{GB}.

\section{2D Electrodynamics}

We sketch the main features of the 2D real time massive
electrodynamics underlying the present work (we denote the mass by
$1/\lambda$ and restrict ourselves to the isotropic case).  We
introduce the metric ${\bf g} = {\rm diag}(-1,-1,+1)$ and define the
gauge field ${\bf a} = ({\bf a}_{xy},a_t)$. In addition, we define the
derivatives $\partial^\alpha = (-{\nabla}_{xy},\partial_t)$, where
${\nabla}_{xy} = (\partial_x, \partial_y)$. The field tensor
$f^{\alpha \beta}$ is given by
\begin{equation}
 f^{\alpha \beta} = \partial^{\alpha} a^{\beta} - \partial^{\beta}
 a^{\alpha} = \left( \begin{array}{ccc} 0 & - b & e_{x} \\ b  &
 0 & e_{y} \\ -e_{x} & -e_{y} & 0 \end{array} \right) 
\label{2D:tensor},
\end{equation}
and its dual takes the form $f^{\ast\alpha} =
\epsilon^{\alpha\beta\gamma} f_{\beta\gamma}= ( e_y, -e_x, -b)$
($\epsilon^{\alpha\beta\gamma}$ is the antisymmetric tensor).  The
Lagrangian of the massive field coupled to an external current
$j^\alpha$ is given by
\begin{equation}
{\cal L} = -\frac{1}{4g^2}f_{\alpha \beta}f^{\alpha \beta} +
\frac{1}{2\lambda^{2} g^{2}}a_\alpha a^\alpha + a_\alpha j^\alpha,
\label{L_aj}
\end{equation}
with $g^2$ the coupling constant.
This Lagrangian produces an imaginary time action
\begin{equation}
{\cal S}[{\bf a}] =  \int \!\! {\rm d} \tau {\rm d}^2 R \left[
 \frac{1}{2g^2}(\nabla\times{\bf a})^2 +
 \frac{1}{2 g^2 \lambda^2}{\bf a}^2 + 
 {\bf a} {\bf j} \right].
\end{equation}
The transformation from real time to imaginary time is obtained via
the formal rules \cite{Zinn} $ t \rightarrow - i \tau$, $\partial_{t}
\rightarrow i \partial_{\tau}$, $a_t \rightarrow -i a_\tau$.  The
magnetic field remains unchanged, while the electric field transforms
to $ {\bf e} \rightarrow i \left[(\nabla\times{\bf
    a})_{xy}\right]_{\perp} = i {\bf e}$.  The free field real time
Lagrangian $\left({\bf e}^2- b^2\right)/2g^2$ then goes over into the
simple form $\left({\bf e}^2 + b^2\right)/2g^2= (\nabla\times{\bf
  a})^2/2g^2$ within the imaginary time formalism.

The functional derivative of Eq.~(\ref{L_aj}) provides us with the
inhomogeneous Maxwell equations (the homogeneous one, $\partial_\alpha
f^{\ast\alpha} = -\partial_t b + {\rm rot} \, {\bf e} = 0$, follows
from the antisymmetric structure of $f^{\alpha \beta}$),
\begin{equation}
\partial_\alpha f^{\alpha \beta} + \frac{1}{\lambda^2}a^\beta 
= g^2 j^\beta.
\label{iME}
\end{equation}
Making use of Gauss' and Stokes' theorems for a loop encircling the
boundary between two media, we obtain the boundary conditions ${\bf
  e}_\parallel = $ continuous and $b = $ continuous.

The quantization of the theory can be developed via the Gupta-Bleuler
formalism within the Lorentz gauge $ \partial_{\alpha} a^{\alpha}=0$.
In the case of a massive theory this method leads to two polarization
modes, while in a massless theory the gauge invariance reduces the
number of polarization modes by one.  Within the present theory, the
coupling of ${\bf a}$ to the real ${\bf A}$ field in the form $i{\bf
  a}\cdot(\nabla\times{\bf A})/\Phi_0$ produces a gauge invariant mass
term ${\bf a}\cdot({\bf a}- {\bf k}\cdot({\bf k}\cdot{\bf
  a})/k^2)/2\lambda^2g^2$.  As the gauge invariance is conserved in
our theory we end up with a single polarization mode.

\end{appendix}



\end{multicols}

\end{document}